\documentclass[10pt,twocolumn]{osajnl}

\journal{ol} 

\setboolean{shortarticle}{false}

\DeclareMathOperator\sgn{sgn}

\title{Supercharge optical arrays}
\author[1,*]{Bikashkali Midya}
\author[2]{Wiktor Walasik}
\author[2]{Natalia M. Litchinitser }
\author[1]{Liang Feng}

\affil[1]{Department of Materials Science and Engineering, University of Pennsylvania, Philadelphia, PA 19104, USA}
\affil[2]{Department of Electrical and Computer Engineering, Duke University, Durham, North Carolina 27708, USA}

\affil[*]{bmidya@seas.upenn.edu}

\begin{abstract}
	We introduce the notion of a {\it supercharge}  optical array synthesized according to supersymmetric charge operators. Starting from an arbitrary array, mathematical supersymmetry transformation can be used systematically to create a zero-energy physical state below the ground state of the super-partner array. This zero mode, which is pinned deep in the midgap of the corresponding supercharge array owing to the square-root spectral relationship between a supercharge and a super-Hamiltonian array, is shown to be  protected because of the chiral symmetry inherent to a supercharge array. A  supercharge array can be used in practical applications to design a discrete optical system of waveguides or coupled resonators where the mid-gap zero mode facilitates robust light dynamics either in spatial or time domain. 
\end{abstract} 

\begin{document}
	
	\maketitle
	
	

	The idea of supersymmetry (SUSY), originally introduced in quantum field theory to unify bosonic and fermionic degrees of freedom~\cite{Wienberg}, became a major mathematical tool in non-relativistic quantum mechanics, statistical, condensed-matter, and optical physics~\cite{Junker,Grover2014,Longhi2010,Miri2013,Bai2006,Longhi2015,Heinrich,Miri2014,Fernandez2014,Principe2015,Ganainy2015,Wiktor2018,Park2015,Macho2018,Midya}.  An underlying supersymmetry of a given  system can be utilized not only to analyze the properties of the system in an elegant 
	 way, but also to design new artificial structures with desirable spectral properties~\cite{Andrianov2004}. In optics, for example, the SUSY technique has been used to synthesize defects that are undetectable by an outside
	observer~\cite{Longhi2010,Midya}, to obtain a transparent interface~\cite{Longhi2015}, and to optimize a family of isospectral quantum cascade lasers~\cite{Bai2006}.  SUSY optical structures have provided a versatile platform to tailor the scattering and localization properties of light, thereby enabling novel applications ranging from phase matching and mode conversion, to spatial multiplexing~\cite{Heinrich,Macho2018}. SUSY has also been extended to transformation optics~\cite{Miri2014}, and to semiconductor laser array design~\cite{Ganainy2015,Wiktor2018}. 
	
	Separate from the development of SUSY, there has been considerable recent interest in optical lattices with zero-energy defect states localized exponentially around the defect, with the corresponding eigenvalues in the gap~\cite{Ge2017,Ge2018,Schomerus2013,Cheng2015,Hadad2017,Biancalana2017,Painter1999,Han2018,Jin2017,Mingsen2018}. Defect modes are capable of achieving subwavelength-scale optical confinement and can be used for enhancement of  nonlinear effects, lasing emission, and cavity quantum electrodynamics. Non-topological defect mode usually has frequency that split from the band edge and is sensitive to disorder---a small structural imperfection can lead to significant resonance frequency detuning. Topological and symmetry protection~\cite{Schomerus2013,Ge2017,Arkinstall2017,Jin2017,Noh2018}, on the other hand, provides robust mode frequency and tighter mode confinement. Nevertheless, starting from scratch, designing an optical lattice with protected defect state remains an elusive task.

\begin{figure}[h!]
		\centering
		\includegraphics[width=0.47\textwidth]{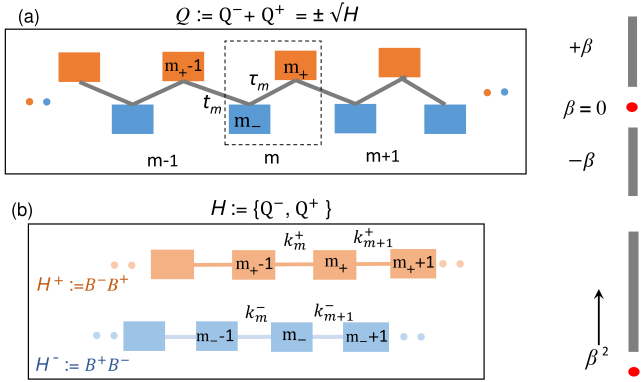} 
		\caption{\small{(a) Schematic of a supercharge array constructed according to bosonic (orange) and fermionic (blue) charge operators. $t_m$ and $\tau_m$ are the `super-hopping' amplitudes between neighboring bosonic and fermionic sites. (b) represents the corresponding super-Hamiltonian with decoupled super-partner arrays. Possible spectrum of these  systems are shown schematically in the right-hand side panels. Midgap zero-energy mode and pairs of positive and negative non-zero modes appear in the supercharge array, while the super-Hamiltonian is degenerate (apart from the zero mode) and is positive definite. See text for the detail relationship between these two systems.}
		}			\label{diagram}
	\end{figure}

	In this letter, we present a novel theoretical framework based on SUSY to synthesize a new class of optical tight-binding models with protected zero-energy modes.  For this aim, we introduce the concept of a {\it supercharge} array. As shown schematically in Fig.~\ref{diagram}, a supercharge array is a binary lattice consisting of interrelated bosonic and fermionic subarrays. The bosonic and fermionic subarrays are characterized by mathematical bosonic and fermionic supercharge operators used to factorize super-partner lattices. (Note that, in optics the terminologies of 'bosonic' and 'fermionic' are rather fictitious; they are used here, in analogy to SUSY quantum theory, for convenient description of the model). The eigen-spectrum of a supercharge array is shown to be related by the square-root that of a super-Hamiltonian. Systematic procedure of SUSY transformations is  utilized to create a mid-gap zero-energy mode at the expense of a defect in the supercharge array; while non-zero modes lie symmetrically in energy around the zero mode. This zero-energy mode can be topologically protected because of the non-zero Witten index and the chiral symmetry of the system. The chiral symmetry, absent in the corresponding super-Hamiltonian system, is a distinctive feature of a supercharge array. 
	
	Another salient feature of a supercharge array is that the bosonic and fermionic subarrays interact with each other; in contrast to a super-Hamiltonian where bosonic and fermionic partner arrays are separable forming a degenerate modes of either type. In the supercharge array, the non-zero energy eigenstates contain both fermionic and bosonic optical modes, residing in fermionic and bosonic sites, respectively;
	while the zero-energy state contains either bosonic or fermioinc mode depending on Witten index $+1$ or $-1$. This, in turn, imply the remarkable opportunity of selective zero-mode enhancement for various applications. For example, in a system with Witten index $+1$, the zero-energy mode can be made to stand out in the bosonic sites provided fermionic sites are complemented by added loss in order to suppress all other non-zero modes residing in both type of sites. This can potentially regulate a single mode and dynamically stable light emission in a resonator network, the challenge for which a solution is being sought presently~\cite{Feng2018}. Nevertheless, as exemplified below, a supercharge waveguide array provides robust light propagation in a confined domain owing to the protected zero mode. 
	
	To introduce the theoretical underpinning of the supercharge array, we briefly discuss the mathematical basis of SUSY theory in quantum mechanics. A quantum mechanical system characterized by a self-adjoint Hamiltonian $H$,
	acting on some Hilbert space $\mathcal{H}$, is supersymmetric if there exist
	supercharge operators $Q^+$ and  $Q^-$ obeying the super-algebra~\cite{Junker}:~$\{Q^+,Q^-\}=H$, and $[Q^\pm,H]=0$. The superscripts `$+$' and `$-$', respectively, refer to `bosonic' and `fermionic', whose relevance will be clear below.
	The supercharges are constructed  as a combination of boson and fermion creation and annihilation operators:~$Q^+ = B^+\otimes f$, $Q^- = (Q^+)^\dag$, where $B^-=(B^+)^\dag$, and $f=(\sigma_x +i \sigma _y)/2$ is expressed in terms of Pauli matrices. When expressed in terms of charge operators, the underlying super-Hamiltonian reduces to $H = H^- \oplus H^+$, with $H^+=B^-B^+$ and $H^-=B^+B^-$ acting on the Hilbert spaces $\mathcal{H}^+$ and $\mathcal{H}^-$, respectively, such that $\mathcal{H}=\mathcal{H}^-\oplus\mathcal{H}^+$. The charge operator $Q^-:\mathcal{H} \rightarrow \mathcal{H}$ transforms a fermionic multiplet, $(|\psi^-\rangle,0)^T$, to a bosonic one, $(0,|\psi^+\rangle)^T$, without changing the energy. Whereas its adjoint, $Q^+$, transforms a  bosonic state to a fermionic one. Although, the supercharges $Q^\pm$ are not self-adjoint, the total supercharge, $\mathcal{Q}$, defined by 
	\begin{equation}\label{Eq-2}
	\mathcal{Q} = Q^+ + Q^- = \left(\begin{array}{cc}
	0 & B^+ \\ 
	B^- & 0 
	\end{array} \right) = \pm \sqrt{H},
	\end{equation}
	is a Hermitian operator i.e. $\mathcal{Q} = \mathcal{Q}^\dag$. This implies that the operator  $\mathcal{Q}$ can be considered for a Hamiltonian description of a physical system. Consequently, the total supercharge $\mathcal{Q}$ is the primary object of investigation in the following. 
	
	Few remarks are in order here. The second equality in \eqref{Eq-2}, which is easy to verify by noting that $\mathcal{Q}^2=diag(B^+B^-,B^-B^+)=H$, implies that the spectrum of $\mathcal{Q}$ is completely determined by that of $H$. If the Schr\"odinger equations for $H^\pm$ satisfy
	$H^\pm |\psi^\pm\rangle = \beta^2 |\psi^\pm\rangle$,
	then the super-multiplet~$|\Psi\rangle =\left(|\psi^-\rangle,|\psi^+\rangle\right)^T$ satisfies 
	\begin{equation} \label{Eq-4}
	H |\Psi\rangle = \beta^2 |\Psi\rangle, ~~~~~~ \mathcal{Q}|\Psi\rangle = \pm \beta|\Psi\rangle,
	\end{equation}
	which shows the appearance of positive and negative energy pairs in the spectrum of $\mathcal{Q}$. This can be explained in terms of an additional symmetry---the chiral symmetry (absent in $H$): $\sigma_z \mathcal{Q} \sigma_z^{-1} = - \mathcal{Q}$ implying  $|\Psi({\beta})\rangle = \sigma_z |\Psi({-\beta})\rangle$. It is possible however for an eigenstate to be its own pair for $\beta = 0$, in this case the state can be protected. Similar situation occurs in topological Jackiew-Rebbi~\cite{Jackiew-Rebbi} and Su-Schrieffer-Heeger (SSH)~\cite{SSH} models where zero-energy state is protected by the domain wall topology of the corresponding scalar field and by the sublattice symmetry, respectively. The emergence of  chiral-symmetry protected zero-energy state, in the supercharge system, is also related to the unbroken symmetry of the corresponding super-Hamiltonian. This can be characterized by the non-zero Witten index~\cite{Witten}. The Witten index, for a system with discrete spectrum, also corresponds to topology of the Hilbert space $\mathcal{H}$ by the index theorem \cite{Alvarez1983}, is defined as: 
	\begin{eqnarray}
	\Delta = Tr \left[(-1)^F e^{-H}\right] = n^+(\beta=0)-n^-(\beta=0),
	\end{eqnarray}
	which is non-zero (zero) for unbroken (broken) symmetry. Here, $n^\pm$ denote the number of zero-energy solutions in the spectrum of $H^\pm$, respectively. The non-zero $\Delta$, therefore, implies the existence of zero mode in the system.  Note that for a system with both discrete and continuous spectrum the above definition leads to anomalous behavior and fails to characterize (broken vs. unbroken) symmetry, and a different approach is necessary~\cite{Akhoury1984}. For the example provided below we consider quantized Zak phase as a topological measure of the zero mode~\cite{Asboth2016}.

	Having set the conceptual framework of a supercharge system, here we show how to construct a supercharge tight-binding lattice. We start from a one-dimensional photonic network, e.g. waveguide arrays or cavities, described by tight-binding Hamiltonian in a form convenient for following discussions (Fig.~\ref{diagram}):
{\small	\begin{equation}
	H^- | m_- \rangle = k_{m+1}^- | m_- +1 \rangle +  k_{m}^- | m_- -1 \rangle  + V_{m}^- | m_-\rangle = \beta^2 |m_-\rangle 
	\end{equation} }
	where $m_-\pm 1 \equiv (m\pm1)_-$, $k_m^-$ is the  nearest-neighbor hopping amplitude between adjacent sites at $m_--1$ and $m_-$, and $V_m^-$ is on-site potential. The eigenvalues, $\beta^2$, correspond to either propagation constants (for waveguides) or resonance frequencies (for cavities). Here, $|m_-\rangle$ denotes Wannier basis vectors localized at site $m_-$, such that $\langle m_-|n_-\rangle = \delta_{mn}, m,n =1,2,...M$. The systems is finite or semi-infinite depending on whether $M$ is finite of infinite. We represent the stationary states of the system as $|\psi^-\rangle = \sum\limits_m \psi_m^{-} |m_-\rangle \in \mathcal{H}^-$, such that modal amplitudes $\{\psi_m^-\}$ satisfy the discrete Schr\"odinger equation
	\begin{equation}\label{H-}
	k_{m+1}^-\psi_{m+1}^- + k_m^- \psi_{m-1}^- + V_m^- \psi^-_m = \beta^2 \psi_m^-.
	\end{equation}
	The state $|\psi^-\rangle$ describes a physically acceptable solution if the following boundary conditions are satisfied: $\psi^-_m = 0$ when $m =0$ and $M+1$, i.e. the energy flow is  prohibited outside the discrete array. A pair of photonic lattices $(H^-,H^+)$, where bosonic lattice $H^+$ have the same form as $H^-$ with `$-$' sign is replaced by `$+$', can be factorized $H^- - \epsilon = B^+ B^-, H^+ -\epsilon=  B^- B^+$ in terms of discrete SUSY transformations~\cite{Matveev}
	\begin{equation}\begin{array}{ll}
	B^- |m_-\rangle  = \tau_m |m_+\rangle + t_m |m_+ - 1\rangle ,\\
	B^+ |m_+\rangle  = t_{m+1} |m_- + 1\rangle + \tau_m |m_-\rangle,
	\end{array}\label{charges}
	\end{equation}
	such that 
	\begin{equation}\label{tm}
	t_m = \sqrt{- k_m^- f_{m-1}^-\big/f_{m}^-},  ~ \tau_m = k_{m+1}^-\big/t_{m+1},
	\end{equation}
	where $f_m^-$ satisfies \eqref{H-} with $\beta^2$ is replaced by the factorization energy $\epsilon$. 
	
	The factorization of two lattices $H^\pm$, implies that starting from $H^-$ with given hopping amplitudes and on-site potentials, it is possible to generate a new Hamiltonian $H^+$ such that $k_m^+ = (k_{m+1}^- t_m)/t_{m+1}, V_m^+ = V_m^- -t_m^2 + t_{m+1}^2$, and corresponding degenerate states (apart from a normalization factor) are related by the charge operators: $|\psi^+(\beta)\rangle = B^- |\psi^-(\beta)\rangle$ and $|\psi^-(\beta)\rangle = B^+ |\psi^+(\beta)\rangle$ such that the mode amplitudes satisfy
	\begin{equation}\label{Eq-SCLattice}
	 \psi_m^+ =  t_{m+1} \psi_{m+1}^- + \tau_m \psi_m^-, ~~~ \psi_m^- = t_{m} \psi_{m-1}^+ +\tau_m \psi_m^+
	\end{equation}
	except for the factorization energy $\beta^2 =\epsilon$. In the later case, the solutions are given by $B^\pm |\psi^\pm(\epsilon)\rangle = 0$, which read
	\begin{equation}\label{Eq8}
	\psi_m^+ = (-1)^{m+1/2}\sqrt{1\Big/\left(k_{m+1}^-f_{m+1}^- f_m^-\right)},~~~\psi_m^- = f_m^-,
	\end{equation}
	respectively. Equation (\ref{Eq8}) implies that both $\psi_m^\pm$ can not both correspond to physical solutions at the energy $\epsilon$, due to the following reason. If $\epsilon$ is less than the ground-state energy of $H^-$, and corresponds to a non-physical solution meaning that $|f_m^-| \rightarrow \infty$ as $m$ approaches boundaries of the network, then $\psi_m^+$ is bounded favoring the state $|\psi^+(\epsilon)\rangle=\sum_m \psi_m^+ |m_+\rangle$ to be normalizable. Consequently, the energy $\epsilon$ can be included in the point spectrum of $H^+$ such that $spec(H^+)=spec(H^-)\bigcup\{\epsilon\}$. Since, in the above factorization procedure, $\epsilon$ was subtracted from $H^\pm$, the super-Hamiltonian is therefore given by $H=diag(H^--\epsilon,H^+-\epsilon)$. As a result, $spec(H)=\{\beta^2-\epsilon\}\bigcup\{0\}$ and corresponding Witten index is given by $+1$. 
	
	\begin{figure}[t!]
		\centering
		\includegraphics[width=0.48\textwidth]{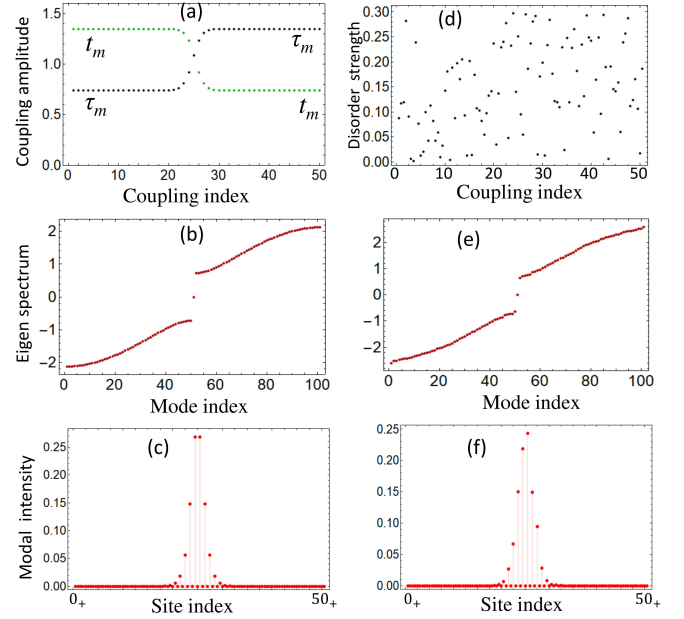} 
		\caption{\small{Example of a topological supercharge array.  (a) shows the supercharge hopping amplitudes for an array with fifty unit cells, which are obtained from a homogeneous parent lattice with $k_m^-=1, V_m^-=0$. (b) and (c) show the spectrum, and zero-mode intensity distribution $(|\psi_m^+|^2,|\psi_m^-|^2)$, respectively. (e) and (f) show same as in (b) and (c) when random hopping disorder are added into the system. Disorder strengths are shown in (d). Clearly, in this case, the zero energy mode remains undisturbed, although rest of the spectrum is perturbed; whereas zero-mode intensity becomes slightly asymmetric due to the strong disorder, but localization width remains intact. Here, $M=50,a=0.6$ and $m_0=25$ are considered.}}\label{fig-2}
	\end{figure}
	
	The above analysis shows that a supercharge array, corresponding to a pair of super-partner photonic lattices, can be constructed  either by Eq.~(\ref{Eq-2}) and (\ref{charges}) or equivalently by Eq.~(\ref{Eq-SCLattice}). Equation (\ref{Eq-SCLattice}) implies that the supercharge array is a binary lattice with each unit cells containing one fermionic and another bosonic sites with mode amplitudes $\psi^-_m$ and $\psi^+_m$, respectively [Fig.~\ref{diagram}(a)]. The couplings, in this case, are given by "super-hopping" amplitudes (in analogy with "super-potential" of continuous case) $t_m$ and $\tau_m$; whereas onsite potential is zero in all the sites. Physically acceptable criterion of $t_m$ (i.e. real valued and free from singularity) can be fulfilled by proper choice of the factorization solution satisfying $f_m^- \ne 0$, and $\sgn(f_m) = (-1)^m$ provided $k_m^- >0$. The spectrum of the supercharge array is obtainable from that of $H$, and is given by $spec(\mathcal{Q})=\{\pm\sqrt{\beta^2-\epsilon}\}\bigcup\{0\}$. Note that, since the fermionic lattice, $H^-$, have no zero-energy counter part, the supercharge zero mode is given by $|\Psi(0)\rangle = (0,\sum_m \psi_m^+ |m_+\rangle)^T$, with non-vanishing bosonic modal amplitudes given by \eqref{Eq8}.
	
\begin{figure}[h!]
		\centering
		\includegraphics[width=0.48\textwidth]{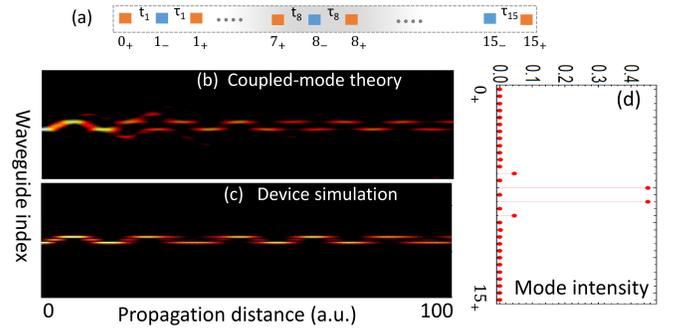} 
		\caption{\small{(a) shows the schematic design of a supercharge array with 31 evanescently coupled  waveguides with identical geometry (only the cross-section of the array is shown). The shaded region implies a defect in the array. (b) and (c) show the intensity evolution, $(|\psi_m^+(z)|^2,|\psi_m^-(z)|^2)$, according to the coupled-mode theory and photonic device simulations (using Comsol multi-physics) when a single site (at $8_+$) is excited in the array. (d) Shows the stationary zero-mode intensity distribution. Here $M=15, a=1.5$ and $m_0=8$ are chosen.}
		}\label{fig-3}
	\end{figure}

Experimental design of a supercharge array, with protected zero mode, can be done by the following recipe: we start from an arbitrary parent lattice $H^-$, with $M$ sites, for which the coupling constants, onsite potentials and the spectrum are known. As a next step, we choose the factorization energy $\epsilon$, below the ground state of $H^-$, such that corresponding non-physical solutions $\{f_m^-(\epsilon)\}$ satisfy  Eq.~(\ref{H-}) and the condition mentioned in the above paragraph.  A supercharge array, then, can be constructed according to Eq.~(\ref{Eq-SCLattice}), with coupling amplitudes given by Eq.~(\ref{tm}). The total number of sites in a supercharge array is $2M+1$, $M$ fermionic and $M+1$ bosonic; an extra bosonic site is due to the fact that the bosonic Hamiltonian has an extra state (the zero mode) than its fermionic counter part. This implies that the boundary (first and last) sites in a supercharge array are of bosonic type, and the fermionic sites are placed in between two consecutive bosonic sites [Fig.~\ref{fig-3}(a)]. The spectrum of a supercharge array can be analyzed conveniently and independently of $H$ if we rewrite $\mathcal{Q}$ in a matrix form according to Eq.~(\ref{Eq-SCLattice}), i.e., $\mathcal{Q}$ is a tridiagonal matrix with main diagonal entries are all zero and off-diagonal elements are $(t_1,\tau_1,\cdots,t_{M},\tau_{M})$. In this representation, supercharge modal amplitudes are given by $|\Psi\rangle=(\psi_0^+,\psi^-_1,\psi^+_1,\cdots,\psi_M^-,\psi^+_M)^T$. 

In particular, when the parent lattice $H^-$ is homogeneous with zero onsite potential i.e. $k_m^-=1$ and $V_m^-=0$ for all $m$, the discrete spectrum can be calculated analytically: $\beta_\ell^2~=~2\cos~\ell \pi/(M+1)$, and $\psi_m^-(\beta_\ell)~=~sin~\ell m\pi/(M+1), 1\le \ell\le M$. In the limit of semi-infinite lattice, the spectrum forms a continuous band in $-2<\beta^2<2$. Such a lattice has been investigated earlier in regard to reflection-less optical structures~\cite{Longhi2010}. By considering factorization energy $\epsilon =-2\cosh~a <\beta_1^2$, at which the unphysical solution is given by $f_m^-=(-1)^m \cosh~a(m-m_0)~$, where $a$ and $m_0$ are two arbitrary real constants, we construct a supercharge array whose coupling distributions are shown in Fig.~\ref{fig-2}(a). The coupling distribution implies that the supercharge array has periodicity defect, and both $t_m$ and $\tau_m$ are mirror anti-symmetric with respect to the defect. Remarkably, away from the defect the inter- and intra-cell couplings appear to be homogeneous; and alternatively strong and weak, respectively, in one side, and the opposite arrangement in other side (similar coupling arrangements were shown to be necessary for  quantization of the Zak-phase in paradigmatic SSH topological chain~\cite{SSH}). It is straight forward to calculate the topological invariant---the Zak phase---in this case by considering left- and right-side periodic bulk lattices (i.e. sufficiently away from the defect where $t_m \rightarrow t$ and $\tau_m\rightarrow\tau$). The Zak phases for the Bloch bands are given by $\pi$ or $0$ depending on $t> \tau$ or $t<\tau$~\cite{Asboth2016}. This justifies the appearance of zero energy defect modes, as shown in Fig.~\ref{fig-2}(b) and (c), at the interface of two topologically different lattices. Remarkable stability of the zero mode in the spectrum is verified by adding up to 30\% random hopping disorder; numerically calculated corresponding spectrum and localized mode amplitudes are shown in Fig.~\ref{fig-2}(e) and (f). Note  that, in this example, the energy gap and position of the defect (hence the center of zero-mode localization) can be controlled by choosing $\epsilon$ (which depends on arbitrary constant $a$) and $m_0$, respectively.

	Finally, we investigate the dynamics of the zero-mode propagation in a supercharge waveguide array, where coupling amplitudes can be manipulated by adjusting the center to center distance between two adjacent waveguides~\cite{Mingsen2018}. Light-intensity evolution in such a system is governed by the coupled-mode equations~\cite{Longhi2009}:~$i\frac{\partial |\Psi\rangle}{\partial z} + \mathcal{Q} |\Psi\rangle=0$, where $z$ is the propagation distance along the waveguide direction. In an array with $31$ waveguides, the propagation dynamics of a single site excitation with moderate random coupling noise is shown in Fig.~\ref{fig-3}(b), and compared in (c) with the direct simulation for silicon ridge coupled waveguides on silica substrate at free space wavelength $1.55\mu$m. Dispersion-less and robust light propagation is observed in this case due to the zero-energy localized mode.
	
	In summary, we have shown that the algebra of a supersymmetric field theory provides an elegant and versatile platform to synthesize a new class of optical arrays---the supercharge arrays characterized by the bosonic and fermionic charge operators. The particular example, shown here, reveals that starting form an arbitrary homogeneous lattice, the mathematical SUSY transformation allows to create a topological defect state in the supercharge array. The supercharge array is shown to be chiral symmetric by construction. By close analogy to the Dirac operator in continuous case, the supercharge array introduced here can be a suitable candidate to emulate relativistic phenomena in discretized photonic systems (supercharge boson-fermion pair in this case is equivalent to a Dirac spinor). Extension of the theory to higher dimensions and its applications to a wide
	range of physical systems as well as the interplay between SUSY breaking and topological phase transition is future direction of investigation.
	
	\noindent {\bf Funding.} Army Research Office (W911NF-17-1-0400).


	\pagebreak
	\makeatother
	\onecolumn

\end{document}